\documentclass{ptapap}
\usepackage{amssymb}
\usepackage{url}
\usepackage{gensymb}
\usepackage{hyperref}
\hypersetup{colorlinks=true,citecolor=blue}
\newcommand{\WISC}{WISE$\times$SCOS}

\author{Maciej Bilicki}[LU,ZG,UCT]

\affil[LU]{Leiden Observatory, Leiden University, The Netherlands}
\affil[ZG]{Janusz Gil Institute of Astronomy, University of Zielona G\'{o}ra, Poland}
\affil[UCT]{Department of Astronomy, University of Cape Town, South Africa}

\title{Cosmology with all-sky surveys}

\begin{document}

\maketitle

\begin{abstract}

Various aspects of cosmology require comprehensive all-sky mapping of the cosmic web to considerable depths. In order to probe the whole extragalactic sky beyond 100 Mpc, one must draw on multiwavelength datasets and state-of-the-art photometric redshift techniques. Here I summarize our dedicated program that employs the largest photometric all-sky surveys -- 2MASS, WISE and SuperCOSMOS -- to obtain accurate redshift estimates of millions of galaxies. The first outcome of these efforts -- the 2MASS Photometric Redshift catalog (2MPZ) -- was publicly released in 2013 and includes almost 1 million galaxies with a median redshift of $z\sim0.1$. I discuss how this catalog was constructed and how it is being used for various cosmological tests. I also present how combining the WISE mid-infrared survey with SuperCOSMOS optical data allowed us to push to depths over 1 Gpc on unprecedented angular scales. These photometric redshift samples, with about 20 million sources in total, provide access to volumes large enough to study observationally the Copernican Principle of universal homogeneity and isotropy, as well as to probe various aspects of dark energy and dark matter through cross-correlations with other data such as the cosmic microwave or gamma-ray backgrounds. Last but not least, they constitute a test-bed for forthcoming wide-angle multi-million galaxy samples expected from such instruments as the SKA, Euclid, or LSST.

\end{abstract}

\section{Introduction}

According to the gravitational instability paradigm, the large-scale structure (LSS) of the Universe has formed over the cosmic time from tiny primordial density fluctuations imprinted today in the cosmic microwave background (CMB) radiation. Galaxies are organized into a network of interconnected filaments and walls, surrounding giant voids, which is known as the \textit{cosmic web}. Observing this web comprehensively requires us to gather representative samples of the Universe: covering large areas of the sky and reaching as deep in redshift as possible. To date, the most successful in this has been the Sloan Digital Sky Survey (SDSS), collecting spectra on $\sim25\%$ of the sky. However, even within such an immense survey, a trade-off between how much of the sky is covered and how deep it can reach must be made: observing the wide-angle three-dimensional (3D) galaxy distribution is expensive and time-consuming.

At the same time, various cosmological probes require \textit{all-sky}, i.e.\ $\lesssim 4 \pi$ sterad, galaxy samples, and preferably in 3D. This is needed to obtain a complete picture of the Universe and to test some of the fundamental assumptions of the standard cosmological model. For instance, the CMB tells as that the early Universe was very homogeneous and isotropic, but the validity of he Copernican Principle (CP) \textit{today} still needs to be studied observationally. Some of the question that arise include: are the CMB anomalies, detected first by WMAP and then confirmed by Planck, imprinted as today's anisotropy and/or inhomogeneity in galaxy distribution? How large are the bulk flows of galaxies, are they in conflict with the CP? What structures pull the Local Group of galaxies? These issues cannot be fully resolved without deep all-sky 3D probes. In addition, many other cosmological tests, such as the integrated Sachs-Wolfe effect (ISW), gravitational lensing of the CMB on LSS or the baryonic acoustic oscillations, benefit from surveys of as large angular coverage and volume as possible.

Several observational programs have mapped the entire sky, both from the ground and space; with a few exceptions, these were mostly photometric surveys. The Two Micron All Sky Survey Extended Source Catalog (2MASS XSC, \citealt{2MASS,2MASS.XSC}) is still the largest all-sky catalog of confirmed extragalactic extended sources\footnote{Note however that extragalactic sources can also be  extracted from the 2MASS \textit{Point} Source Catalog, as shown by \cite{KoSz15} and confirmed by \cite{Rahman2MASS}.}, containing 1.6 million galaxies with $JHK_s$ photometric information and completeness limit of $K_s\sim13.9$ (Vega), equivalent to $\langle z \rangle \sim 0.1$. Its $K_s<11.75$ flux-limited subset, 2MASS Redshift Survey of 44,000 galaxies (2MRS, \citealt{2MRS}), is until now the biggest complete spectroscopic dataset covering the whole sky, despite peaking only at $\langle z \rangle = 0.03$. 

There do exist  all-sky samples much deeper than 2MASS, provided by the Wide-field Infrared Survey Explorer (WISE, \citealt{WISE}) as well as the SuperCOSMOS scans of photographic plates (SCOS, \citealt{SCOS}). WISE satellite data include an all-sky photometric catalog at 3.4, 4.6, 12 and 23 $\mu$m, with almost 750 million detections in its `AllWISE' release; however, only the two former bands are uniformly measured to the full depth of the survey (about 17 Vega mag in the 3.4 $\mu$m channel). SCOS includes photometric and morphological information of almost 1 billion objects in the $B,R,I$ passbands, only a fraction of which are however reliable detections useful for constructing uniform all-sky galaxy catalogs. In general, the two datasets are on their own limited by the lack of unambiguous source type identification; both also  severely suffer from star blending at low Galactic latitudes.

In the present proceedings I summarize how using the above-listed datasets \textit{in concert} -- in the spirit of modern multi-wavelength astrophysics -- allowed us to produce value-added catalogs which are now being used for various cosmological applications, many of them inaccessible beforehand.

\section{2MPZ and its cosmological applications}

\begin{figure}
\centering
\includegraphics[width=\textwidth]{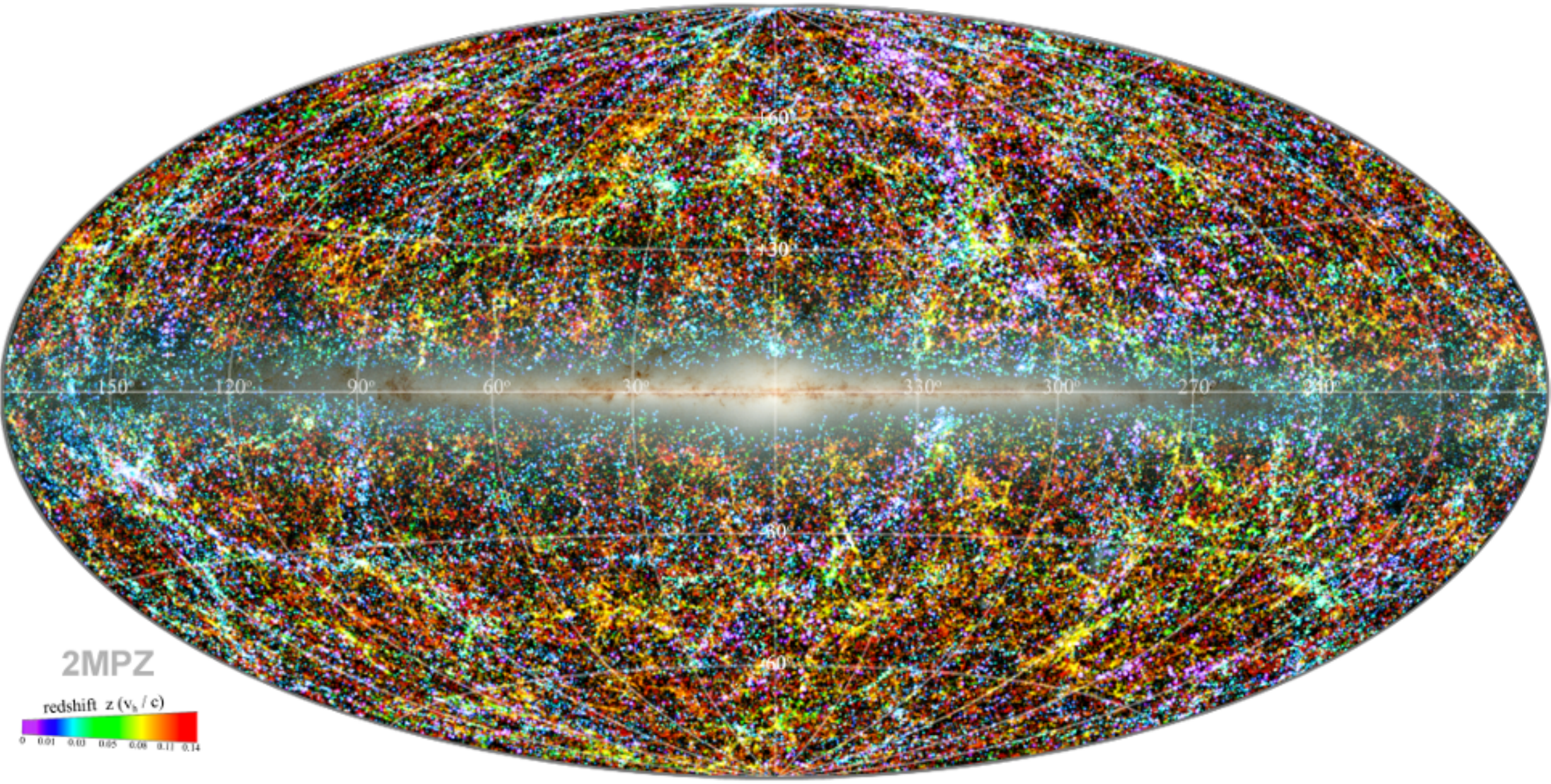} 
\caption{\label{Fig:2MPZ}Distribution of 2MPZ galaxies in Aitoff projection in Galactic coordinates, color-coded by photometric redshift. Plot courtesy of Tom Jarrett.}
\end{figure}

At present, about 1/3 of all 2MASS galaxies have spectroscopic redshifts (spec-$z$) from different wide-angle surveys (see 2M++ by \citealt{2M++}). A possibility is now emerging that the remaining 2/3 will be supplemented with spec-$z$ as well, thanks to the starting TAIPAN\footnote{\url{http://www.taipan-survey.org/}} and proposed LoRCA \citep{LORCA} surveys. However, before this happens, the only way to add the 3rd dimension to 2MASS at its full depth is through \textit{photometric redshifts} (photo-$z$). This was done earlier by \cite{XSCz} using only 2MASS data, and by \cite{FP10} who added also SCOS. Thanks to WISE all-sky information, released in 2012,  it was possible to gather 8-band photometry (from $B$ to 4.6 $\mu$m) for the majority ($\sim95\%$) of 2MASS extended sources. By training the ANN$z$ algorithm \citep{ANNz} on the spec-$z$ subsample, we derived unbiased and precise photo-$z$s for these objects, with a scatter of $\sigma_{\delta z} \sim 0.013$ (see \citealt{2MPZ} for details). The resulting 2MASS Photometric Redshift sample (2MPZ) of almost 1 million galaxies covering most of the sky is publicly available\footnote{\url{http://ssa.roe.ac.uk/TWOMPZ.html}} and is illustrated in Fig.\ \ref{Fig:2MPZ}.

2MPZ has largely contributed to the recent revival of interest in using 2MASS data for cosmological applications, some of which with the involvement of the present author. For instance, the Planck team used 2MPZ to reconstruct the gravitational potential in the context of their ISW analysis \citep{PlanckXXI}. \cite{AS14} tested global isotropy employing this dataset, finding no deviations therefrom in the local Universe. \cite{Alonso15} showed that 2MPZ galaxy distribution is inconsistent with simple fractal models. Very recently, \cite{KGB15} proposed an explanation of the CMB `cold spot' as an imprint of a chain of elongated voids identified in 2MPZ and 6dFGS data. In a forthcoming paper, we show that 2MPZ is too shallow to provide an all-sky detection of the ISW, even through a tomographic analysis, once systematics are accounted for ({\color{blue}Steward et al., in prep.}). The catalog is now also used for an angular correlation analysis in redshift shells, as a test-bed for future surveys such as Euclid ({\color{blue}Balaguera-Antol{\'{\i}}nez et al., in prep.}). I expect that 2MPZ will find many more applications in the coming years, as it will remain the most comprehensive all-sky 3D census of $z\sim0.1$ galaxies until all the 2MASS galaxies are measured spectroscopically.

\section{WISE$\times$SuperCOSMOS}

The depth of 2MPZ is limited by the shallowest of the parent surveys, 2MASS. It was then an obvious exercise to cross-match WISE with only SCOS and generate photo-$z$s for the resulting sample. A proof of concept was provided already in \cite{2MPZ} and the final results are now described in detail in \cite{WISC}. Here it was not possible to cover as much of the sky as in 2MPZ due to severe stellar blending in the two component catalogs, which mimics extended sources, as well as because of extinction blocking optical light; still, about $3\pi$ sterad are reliably covered by the sample. Stellar contamination is in general an issue here, due to the limited quality of SCOS data and a large, $6"$, beam of WISE. Out of 170 million sources from a cross-match of \textit{extended} SCOS objects with WISE at $|b|>10\degree$ within fiducial flux limits,
 only about 20 million are indeed galaxies.  This number was obtained in \cite{WISC} from an adaptive, WISE-based color cut, but is also confirmed in an independent machine-learning analysis by {\color{blue}Krakowski et al.\ (in prep.)} 
(see also {\color{blue}Ma{\l}ek et al.}\ in this volume).

\begin{figure}
  \centering
  \begin{minipage}{0.48\textwidth}
\includegraphics[width=\textwidth]{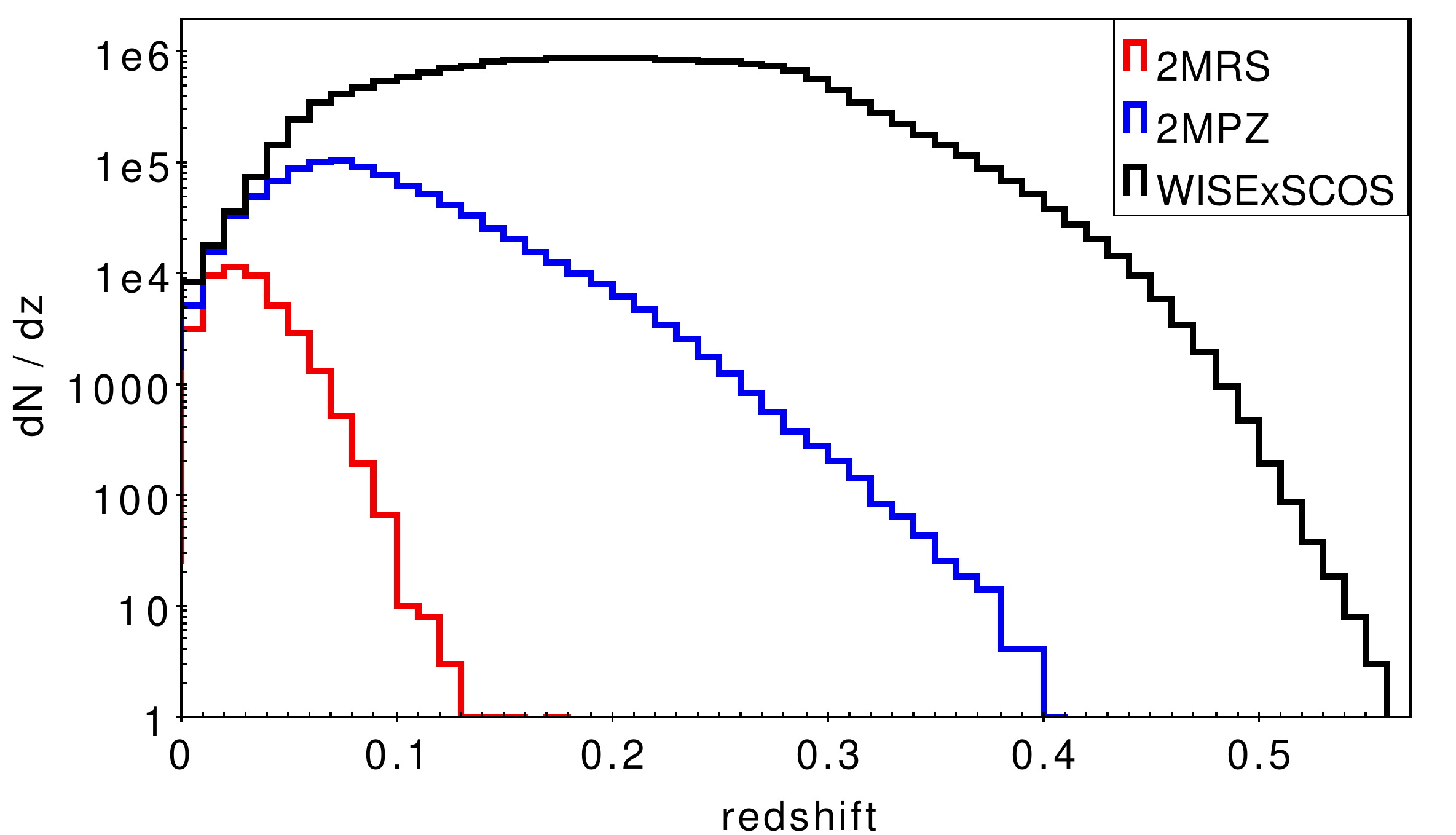} 
\caption{\label{Fig:dNdzs}Redshift distributions of three all-sky datasets: 2MRS, 2MPZ, and WISE$\times$SuperCOSMOS.}
  \end{minipage}
  \quad
  \begin{minipage}{0.48\textwidth}
\includegraphics[width=\textwidth]{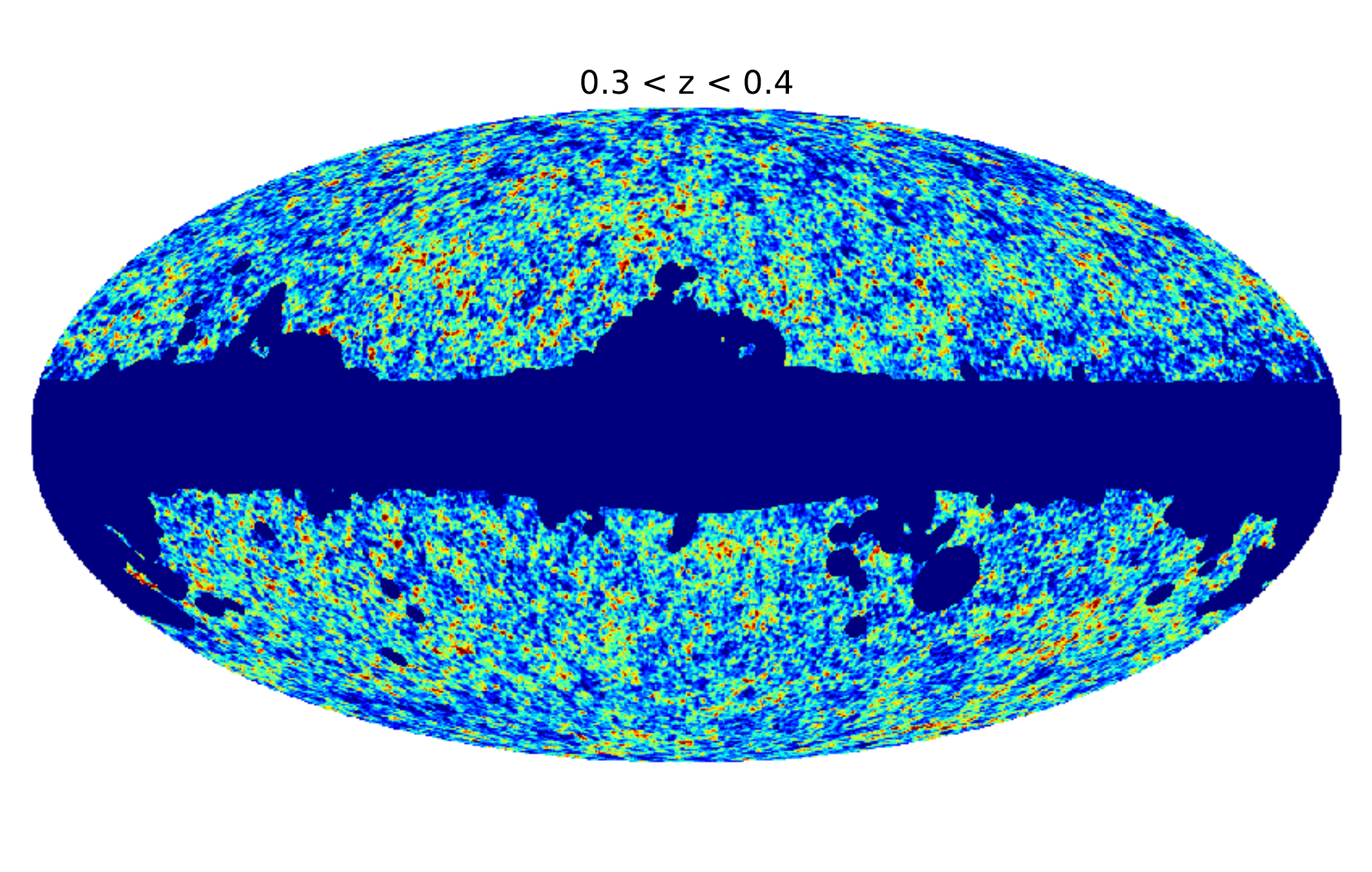} 
\caption{\label{Fig:WISCmap}Pixelized map of WISE$\times$SuperCOSMOS galaxy distribution in the $0.3<z_{\mathrm{phot}}<0.4$ shell. Dark blue is the mask applied on the data.}
  \end{minipage}
\end{figure}

Despite only 4-band photometry available for the full \WISC\ sample ($B$, $R$, 3.5 \& 4.6 $\mu$m), we were able to derive reliable photo-$z$s for all the 20 million galaxies, with overall $\sigma_{\delta z}= 0.033$ and other statistics comparable to the 2MPZ case. This was done using also the ANN$z$ algorithm, and was possible thanks to the availability of excellent GAMA-II spectroscopic data for calibration, very complete to $r<19.8$ in 3 equatorial fields \citep{Liske15}. The median redshift of the resulting sample is $z\sim0.2$, but its relatively flat redshift distribution probes the LSS up to $z\sim0.4$ (Fig.\ \ref{Fig:dNdzs}). At these redshifts, \WISC\ is the only photo-$z$ dataset with such a sky coverage, which makes it suitable for various applications that benefit from these properties. Fig.\ \ref{Fig:WISCmap} gives an example of a map in the photo-$z$ shell of $0.3<z<0.4$, illustrating the LSS roughly 4 billion years ago.

We plan to release the final \WISC\ catalog in 2016, hoping that it will find at least as many uses as 2MPZ has already had. In the meantime, some of its applications are already ongoing. {\color{blue}Peacock \& Bilicki (in prep.)} will present the results of a tomographic cross-correlation of the dataset with the CMB lensing convergence field measured by Planck, to provide constraints on the redshift evolution of the growth of structure. In {\color{blue}Cuoco et al.\ (in prep.)}, Fermi-LAT $\gamma$-ray background will be cross-correlated with \WISC\ in redshift shells to look for signals of self-annihilating or decaying dark matter (cf.\ \citealt{Cuoco15}). In general, the catalog should be useful for cosmological applications that require very wide-angle coverage, but are impractical with shallower surveys such as those based on 2MASS. It should also find similar uses as 2MPZ, but at a 3 times larger depth, at a price of smaller sky coverage and more severe systematics. 

\section{All-sky probes: the promise of WISE}

WISE is much deeper than both 2MASS (by $\sim3$ mag) and SCOS. It detects $L^*$ galaxies up to $z\sim0.5$ but probes the LSS possibly to $z\sim1$ or more \citep{JarrettG12}. This should in principle provide another ``layer'' of all-sky data for observational cosmology, beyond 2MPZ and \WISC. However, this cosmological potential is very difficult to take full advantage of, as the WISE data are dominated by stars even at high Galactic latitudes \citep{Jarrett11}. Lack of an extended source catalog in WISE, as well as the availability of only 2 bands at its full depth, make general star/galaxy separation in this dataset a challenging problem. In {\color{blue}Kurcz et al.}\ in this volume we present first steps towards identifying stars, galaxies and QSOs in WISE in an automatized way; more details will be provided in a forthcoming paper ({\color{blue}Kurcz et al., in prep.}).

At the full depth of WISE there is no other all-sky catalog that could be used to supplement the former with reliable photo-$z$s, as was possible in 2MPZ and \WISC. At present, and in the near future, the only possibility to add the 3rd dimension to WISE is by combining it with other datasets, preferably deeper, although covering (much) less of the sky. The examples here are the Vista Hemisphere Survey (near-IR, $2\pi$ sterad), Dark Energy Survey (optical, 5000 deg$^2$) or the Kilo Degree Survey (optical, 1500 deg$^2$). Using such catalogs should provide several additional passbands for WISE sources and lead to such studies as tomographic analyses (based on photo-$z$s) or identification of specific sources, as QSOs, some of which being rare (e.g.\ at very high redshifts).

Last but not least, when working with catalogs such as from WISE, we are getting prepared for forthcoming and future datasets, where objects will be counted in billions rather than millions: LOFAR, SKA, LSST, Euclid... However, even in this `very big data' observational cosmology era, the catalogs presented here, or at least analyses based on them, may be still of use. For instance, both Euclid and especially LSST will not have full spectroscopic redshift coverage, being mostly photometric probes. On the other hand, radio surveys such as LOFAR or SKA will need source identification and/or multiwavelength information from the optical and IR.

\section{Summary}
To summarize the present proceedings, let me first reiterate that all-sky galaxy surveys are essential to comprehensively map the cosmic web. In addition, many key cosmological applications require very wide-angle coverage in 3D. Unfortunately, in the coming years the full extragalactic sky will not be mapped spectroscopically beyond $z\sim0.1$ or so\footnote{This can however change in the longer term thanks to the SPHEREX concept \citep{SPHEREX}.}. Probing the third dimension at such angular scales is currently possible only with photometric redshifts, and thanks to 2MASS, SuperCOSMOS and WISE, we now have access to cosmologically useful volumes over $3\pi$ sterad of the sky. The new `all-sky' galaxy catalogs, presented here, reach up  to $z\sim0.4$, thus opening windows for new cosmological studies.

\acknowledgements{The work presented here was supported by: the Polish National Science Center (NCN) under contract \#UMO-2012/07/D/ST9/02785; the South African National Research Foundation (NRF); the Netherlands Organization for Scientific Research, NWO, through grant \#614.001.451; and the European Research Council through FP7 grant \#279396.}

\bibliographystyle{ptapap}
\bibliography{ptapapdoc}

\end{document}